\documentclass[aps,prd,10pt,onecolumn,final,notitlepage,oneside,nobibnotes,
nofootinbib,superscriptaddress,showkeys,centertags]{revtex4-1}

\usepackage{amsmath,amsfonts,amssymb}
\usepackage{bm}
\usepackage{dcolumn}
\usepackage[dvips]{graphicx,graphics,color,epsfig}
\usepackage{latexsym}
\usepackage{natbib}
\usepackage{verbatim}
\usepackage{bm}
\usepackage{tabulary}
\usepackage{natbib}
\usepackage{mathtext}
\usepackage[T2A]{fontenc}

\def\gccm{g/cm$^{-3}$}
\def\tet{\theta_\mathrm{n}}
\newcommand{\WID}[2]{#1_{\mathrm{#2}}}
\def\No{\textnumero}

\begin{document}
\title{Neutrino Fluxes from a Core-Collapse Supernova in a Model with Three
Sterile Neutrinos}

\author{\firstname{A.~V.}~\surname{Yudin}}
\email{yudin@itep.ru}
\affiliation{Institute for Theoretical and Experimental Physics, Bolshaya
Cheremushkinskaya 25, Moscow, 117218 Russia}
\author{\firstname{D.~K.}~\surname{Nadyozhin}}
\email{nadezhin@itep.ru}
\affiliation{Institute for Theoretical and Experimental Physics, Bolshaya
Cheremushkinskaya 25, Moscow, 117218 Russia}
\author{\firstname{V.~V.}~\surname{Khruschov}}
\email{khruschov\_vv@nrcki.ru}
\affiliation{National Research Center Kurchatov Institute, Academician
Kurchatov Place 1, Moscow, 123182 Russia}
\affiliation{Center for Gravitation and Fundamental Metrology, VNIIMS,
Ozernaya Street 46, Moscow, 119361 Russia}
\author{\firstname{S.~V.}~\surname{Fomichev}}
\email{fomichev\_sv@nrcki.ru}
\affiliation{National Research Center Kurchatov Institute, Academician
Kurchatov Place 1, Moscow, 123182 Russia}
\affiliation{Moscow Institute of Physics and Technology (State University),
Institutskii Lane 9, Dolgoprudnyi, Moscow Region, 141700 Russia}

\begin{abstract}
The characteristics of the gravitational collapse of a supernova and the
fluxes of active and sterile neutrinos produced during the formation of its
protoneutron core have been calculated numerically. The relative yields of
active and sterile neutrinos in core matter with different degrees of
neutronization have been calculated for various input parameters and various
initial conditions. A significant increase in the fraction of sterile
neutrinos produced in superdense core matter at the resonant degree of
neutronization has been confirmed. The contributions of sterile neutrinos to
the collapse dynamics and the total flux of neutrinos produced during collapse
have been shown to be relatively small. The total luminosity of sterile
neutrinos is considerably lower than the luminosity of electron neutrinos, but
their spectrum is considerably harder at high energies.
\end{abstract}

\keywords{active and sterile neutrinos, neutrino oscillations,
neutrino--matter interaction, supernova, collapse}

\maketitle

\section*{INTRODUCTION}
During the gravitational collapse of supernova cores
the bulk of the released energy is carried away by
powerful neutrino fluxes. Neutrinos of various flavors
are produced both through a large number of processes
involving nucleons and nuclei of stellar matter
and due to neutrino oscillations, i.e., the conversion
of one neutrino flavor into another. A peculiar
feature of the oscillations of known neutrino flavors
(electron, muon, and tau) in matter is their
dependence on the electron number density distribution
and the emergence of Mikheev--Smirnov--Wolfenstein (MSW)
resonances. In those regions of
stellar matter where the conditions for the emergence
of MSW resonances are satisfied, the transition of
one neutrino flavor to another is enhanced even if
the initial vacuum mixing between different neutrino
flavors was insignificant. Apart from the oscillations
of known active neutrinos (AcN: electron, muon
tau), the oscillations involving sterile neutrinos (StN)
that do not interact directly with Standard Model
(SM) particles through photon, $W$ and $Z$ boson, and
gluon exchanges can occur.

The StN existence problem is a topical problem of
modern neutrino physics. In particular, it relates
to the anomalies of the neutrino and antineutrino
fluxes at small distances in a number of ground-based
experiments (see Abazajian et al. 2012; Schwetz
et al. 2011; Giunti et al. 2013; Kopp et al. 2013;
Gorbunov 2014). The presence of such anomalies,
if it will be confirmed at a sufficient confidence level,
is beyond the scope of the SM and the minimally
extended SM (MESM) with three active massive
neutrinos, because a new type of neutrinos, namely
sterile neutrinos with a mass scale of $\sim 1$~eV, is required
to explain them. In principle, the StN masses
can lie within a wide range, from $10^{-5}$~eV to $10^{15}$~GeV
(de Gouvea 2005; Drewes and Garbrecht 2015). It
is convenient to divide this range in such a way that
StN with masses less than $0.1$~eV, from $0.1$ to $100$~eV,
from $100$~eV to $10$~GeV, and more than $10$~GeV to be
assigned to the classes of superlight, light, heavy, and
superheavy StN, respectively.

Phenomenological models with one, two, and
three StN have been proposed to take into account
StN (see Abazajian et al. 2012; Schwetz et al. 2011;
Giunti et al. 2013; Kopp et al. 2013; Gorbunov 2014;
Canetti et al. 2013; Conrad et al. 2013). If we take
into account the possible left--right symmetry of weak
interactions and associate the sterile neutrinos with
the right neutrinos neutral with respect to SU(2)$_L$ weak
interactions, then there must be three StN,
i.e., $(3 + 3)$ models with three AcN and three StN
should be considered (Canetti et al. 2013; Conrad
et al. 2013). One of such models is the $(3 + 1 +
2)$ model (or $(3 + 2 + 1)$ model) that contains three
StN, two of which are approximately degenerate in
mass, while the mass of the third one can differ
significantly from the masses of the two other ones
(Zysina et al. 2014; Khruschov and Fomichev 2015).
Within the $(3 + 2 + 1)$ model with three StN the
AcN and StN mass characteristics were estimated
(Zysina et al. 2014), the AcN and StN appearance
and survival probabilities in the Sun were calculated
by taking into account the coherent scattering of
neutrinos in matter (Khruschov and Fomichev 2015),
and various AcN and StN mixing matrices were
considered to explain the neutrino anomalies at small
distances (Khruschov et al. 2016).

Khruschov et al. (2015) considered the change in
neutrino flavor composition due to coherent neutrino
scattering by electrons and neutrons of superdense
matter in the model with three StN. Allowance for
neutrons does not lead to a change in oscillation
characteristics if only AcN are considered. If the
contribution of StN is taken into account, then the
influence of the neutron density of matter becomes
noticeable. In this paper it was shown that when
the neutron-to-proton ratio is close to two, the StN
yield is enhanced considerably. Such an enhancement
arises at high or ultrahigh densities of matter;
therefore, this effect can be of importance
only in astrophysical conditions, for example,
during the formation of a protoneutron supernova
core. In models involving StN this effect is additional
to the MSW effect and can lead to new consequences
during supernova explosions. Despite the fact that
the influence of StN on the processes in supernovae
have been considered in many papers (see, e.g., Hidaka
and Fuller 2006, 2007; Tamborra et al. 2012;
Warren et al. 2014), the effect of an enhancement of
the StN yield at a neutron-to-proton ratio in matter
$\tet\equiv\WID{n}{n}/\WID{n}{p}\approx 2$ was not studied in detail,
although it was pointed out, for example, in Wu et al. (2014).

In models with three AcN and three StN there
are difficulties in reconciling the number of neutrinos
with cosmological observations. The observations of
cosmological cosmic microwave background fluctuations
within the framework of standard cosmological
models limit the number of new light neutrinos
virtually to one (see, e.g., Komatsu et al. 2011;
Ade et al. 2013). Nevertheless, just as in Zysina
et al. (2014), Khruschov and Fomichev (2015), and
Khruschov et al. (2015, 2016), we will consider a
model with three AcN and three StN. Let us list
several reasons for such a choice. These include,
first, the principle of left--right symmetry noted above,
which may be restored on a scale exceeding the scale
of spontaneous electroweak symmetry breaking,
$\sim 1$~TeV. In addition, the form of the AcN and StN
mixing matrix chosen under strict unitarity conditions
in Khruschov et al. (2016) necessitates introducing
three StN. Even if the derived cosmological
constraints on the total number of relativistic neutrinos
are taken into account, it should be noted that
these constraints are model--dependent. Therefore,
there must not be ruled out the possibility of the realization
of nonstandard cosmological
models within which, despite the existence
of three StN, their effect will be reduced or virtually
suppressed in the observational cosmological data, if
other interaction and thermal equilibrium conditions
for StN are assumed (see, e.g., Ho and Scherrer 2013;
Chu et al. 2015).

The goal of this paper is to investigate the generation
of sterile neutrinos during a supernova explosion.
The paper is organized as follows. Initially,
we provide information about the oscillation model
used. Subsequently, we describe the procedures for
calculating the production and propagation of sterile
neutrinos that differ for the case of high opacity and
the semitransparent case. We separately discuss the
neutrino--neutrino scattering effect, which turns out
to be important under conditions of an opaque and
hot protoneutron star forming during a supernova
explosion. We then present the results of our calculations
for two chosen instants of time: at the phase
of collapse and 3 ms after its stop and core bounce.
We calculate the fluxes of sterile neutrinos and their
spectra at the exit from the star. The physical conditions
in the collapsing stellar core obtained in the
hydrodynamic calculations by Yudin (2009), Yudin
and Nadyozhin (2008), and Liebend\"{o}rfer (2005) are
taken into account in this case. In conclusion, we
discuss our results and consider the possibilities of
further studies in this area.

\section*{THE $(3 + 2 + 1)$ MODEL OF ACTIVE AND STERILE NEUTRINOS}
Let us give the basic principles of the phenomenological
$(3 + 2 + 1)$ model of neutrinos that is used
below to calculate the effects involving three StN
(Zysina et al. 2014; Khruschov and Fomichev 2015;
Khruschov et al. 2015, 2016). In this model two
StN are approximately degenerate in mass, while the
mass of the third StN can differ significantly from the
masses of the two other ones. Different StN flavors
are denoted by indices $x$, $y$ and $z$, while the additional
massive states are denoted by indices $1'$, $2'$ and $3'$. The set of
indices $x$, $y$ and $z$ is denoted by symbol $s$, while the set
of indices $1'$, $2'$ and $3'$ is denoted by symbol $i'$. The general
$6\times 6$ mixing matrix $\widetilde{U}$ of AcN and StN can then be
represented via $3\times 3$ matrices $S$, $T$, $V$ and $W$ as
\begin{equation}
\left(\begin{array}{c}
\nu_{\rm a}\\ \nu_{\rm s} \end{array}\right)=
\widetilde{U}\left(\begin{array}{c}
\nu_i\\ \nu_{i'}\end{array}\right)
\equiv
\left(\begin{array}{cc}
S&T\\ V&W\end{array}\right)
\left(\begin{array}{c}
\nu_i\\ \nu_{i'}\end{array}\right).
\label{eq1}
\end{equation}
The neutrino masses are specified in the form $\{m\}=\{m_i,m_{i'}\}$, with
$\{m_i\}$ being arranged in direct order,
i.e., $\{m_1,m_2,m_3\}$ and $\{m_{i'}\}$ being arranged in
reverse order, i.e., $\{m_{3'},m_{2'},m_{1'}\}$. The unitary $6\times 6$ matrix
$\widetilde{U}$ is considered in a certain form under additional
physical assumptions (Khruschov et al. 2016).
As the basis of massive sterile states we choose the
states for which the matrix $W$
is approximately a unit
one while restricting ourselves to a diagonal matrix
of the form $W=\widetilde\varkappa I$, where $I$ is a unit matrix and
$\widetilde\varkappa$ is a complex parameter with a modulus close to
unity, $\widetilde\varkappa=\varkappa\exp{(i\phi)}$. We will represent
the matrix $S$ as $S=U_{PMNS}+\Delta U_{PMNS}$, with the matrices
$\Delta U_{PMNS}$ and $T$ being small compared to $U_{PMNS}$. For the
convenience of our estimations of the mixing
between AcN and StN and the corrections to the mixing
only between AcN due to the influence of StN, we
will assume that
$\Delta U_{PMNS}=-\epsilon U_{PMNS}$, where $\epsilon$ is
a small quantity, with
$\epsilon=1-\varkappa$. The matrix $S$ is then $S=\varkappa U_{PMNS}$, where
$U_{PMNS}$ is the well--known unitary $3\times 3$
Pontekorvo--Maki--Nakagawa--Sakata mixing matrix ($U_{PMNS}U_{PMNS}^+=I$).

Thus, in the model under consideration at an appropriate
normalization AcN are mixed via the matrix
$U_{PMNS}$. At present, the most accurate values of
the neutrino mixing parameters have been obtained
in several papers (see, e.g., Olive et al. 2014; Capozzi
et al. 2014). Since only the absolute value of the
oscillation mass characteristic $\Delta m^2_{31}= m^2_{3}-m^2_{1}$
is
known, the absolute values of the neutrino masses
can be ordered in two ways:  $a)\; m_1<m_2<m_3$ and $b)\; m_3<m_1<m_2$,
i.e., either normal hierarchy (NH,
case a) or inverted hierarchy (IH, case b) of the neutrino
mass spectrum is said to be realized. Given
that the mixing between AcN and StN is small and is
determined by the small parameter $\epsilon$, we will choose
the matrix $T$ in the form$T=\sqrt{1-\varkappa^2}\,a$ where $a$  is an
arbitrary unitary $3\times 3$ matrix, $aa^+=I$. The unitary matrix
$\widetilde{U}$ can then bewritten as
\begin{equation}
\widetilde{U}=
\left(\begin{array}{cc}
\varkappa U_{PMNS}&\sqrt{1-\varkappa^2}\,a\\
-e^{i\phi}\sqrt{1-\varkappa^2}\,a^+U_{PMNS}&\varkappa e^{i\phi}I
\end{array}\right).
\label{eq2}
\end{equation}
Zysina et al. (2014), Khruschov and Fomichev (2015),
and Khruschov et al. (2015) used an approximately
unitary mixing matrix of form (\ref{eq2}) at $\phi=0$, where the
condition for the conservation of the normalization
of StN states was disregarded (for the corrections
related to the nonunitarity of the mixing matrix, see,
e.g., Antusch and Fischer 2014).

In this paper we will choose a trial value of the
unknown phase $\phi$ as $\phi=\pi/4$. The matrix $U_{PMNS}$
also contains an additional CP-phase $\delta_{CP}$. Despite
the fact that the CP-phase has not yet been established
experimentally, it was estimated in several papers
(see, e.g., Capozzi et al. 2014; Khruschov 2013;
Petkov et al. 2015) for the normal hierarchy (NH)
of the neutrino mass spectrum or, more specifically,
$\sin\delta_{CP}<0$ and $\delta_{CP}\approx -\pi/2$. The NH case is also
more preferable when taking into account the constraints
on the sum of the neutrino masses from
cosmological observational data (Huang et al. 2015).
Therefore, in our subsequent numerical calculations
we will assume the NH case to be the main one and
will use $\delta_{CP}=-\pi/2$ for it.

Khruschov et al. (2016) considered three forms
of the matrix $a$ for the NH case: $a_1$, $a_2$, and
$a_3$. It was shown that in the forms $a_1$ and $a_3$
of the mixing matrix the transition probability of
muon neutrinos/antineutrinos to electron neutrinos/
antineutrinos contains no contributions from
StN, and, in both cases, the transition probabilities
are identical and determined only by the AcN mixing
parameters. In this paper we will consider the matrix
$a_2$ written out below as the main one. This form
leads to the dependence of the transition probabilities
of muon neutrinos/antineutrinos to electron
neutrinos/antineutrinos on the StN contributions
and makes it possible to describe the experimentally
revealed neutrino anomalies at small distances in
principle. The matrix $a_2$ is
\begin{equation}
a_{2}=\left(\begin{array}{lcr}
\cos\eta_2 & \sin\eta_2 & 0\\
-\sin\eta_2 & \cos\eta_2 & 0\\
0 & 0 & e^{-i\chi_2}\end{array}\right),
\label{eq3}
\end{equation}
where $\chi_2$ and $\eta_2$ are the phase and angle of mixing
between AcN and StN. In our calculations we will use
the following trial values of the new mixing parameters:
$\chi_2=-\pi/2$ and $\eta_2=\pm 30^{\circ}$. We will constrain
the small parameter  $\epsilon$ by the condition $\epsilon\lesssim 0.03$.

We use the results from Zysina et al. (2014) and
Khruschov and Fomichev (2015) referring to the
estimates of the absolute values of the AcN masses
$m_i$ ($i=1,2,3$) for the NH case in eV: $m_1\approx 0.0016$,
$m_2\approx 0.0088$, $m_3\approx 0.0497$. We choose the
masses $m_{3'}$ and $m_{2'}$ of two StN near $1$~eV in
accordance with the results from Kopp et al. (2013)
for the best StN masses in a $(3 + 2)$ model, i.e.,
$m_{3'}\approx 0.69$~eV and $m_{2'}\approx 0.93$~eV. For the mass $m_{1'}$
of the third StN we will use the mass of a possible dark
matter (DM) particle taken from Bulbul et al. (2014),
Boyarsky et al. (2014), and Horyuchi et al. (2015),
i.e., $m_{1'}\approx 7100$~eV. Thus, the absolute values of the
masses (in eV) of all neutrinos within this model are
as follows:
\begin{equation}
m_{\nu}=\{0.0016,\,0.0088,\,0.0497,\,0.69,\,0.93,\,7100\}.
\label{eq4}
\end{equation}
Given this sterile neutrino mass distribution, the
model with three sterile neutrinos under consideration
may be called the $(3 + 2 + 1)$ model of active and
sterile neutrinos. Note that this model can in principle
mimic the $(3 + 1)$ and $(3 + 2$) models as one or two
sterile neutrinos decrease in importance, for example,
as their masses increase significantly or for a special
choice of mixing angles.

To calculate the probability amplitudes for the
propagation of neutrinos with certain flavors in a
medium, we will use the equations from Khruschov
and Fomichev (2014):
\begin{equation}
i\partial_{r}\left(\begin{array}{c}
a_{\rm a} \\
a_{\rm s}\end{array}\right)
=\left[\frac{\tilde{\Delta}_{m^2}}{2E}
+\sqrt{2}G_{F}\left(\begin{array}{cc}
\widetilde{N}_{\rm e}(r) & 0 \\
0 & \widetilde{N}_{\rm n}(r)/2\end{array}\right)\right]
\left(\begin{array}{c} a_{\rm a}\\
a_{\rm s}\end{array}\right).
\label{ur}
\end{equation}
Here, we introduce the matrix
$\tilde{\Delta}_{m^2}=\tilde{U}\Delta_{m^2}\tilde{U}^T$,
where $\Delta_{m^2}={\rm diag}\{m_{1}^{2}-m_{0}^{2},\,m_{2}^{2}-m_{0}^{2},\,
m_{3}^{2}-m_{0}^{2},\,m_{3'}^{2}-m_{0}^{2},\,m_{2'}^{2}-m_{0}^{2},\,
m_{1'}^{2}-m_{0}^{2}\}$, with $m_0$ being
the smallest neutrino mass among $m_i$ and $m_{i'}$. The
matrices $\widetilde{N}_{\rm e}(r)$ and $\widetilde{N}_{\rm n}(r)$
are $3\times3$ matrices of the
form
\begin{equation}
\widetilde{N}_{\rm e}(r)=\left(\begin{array}{ccc}
n_{\rm e}(r) & 0 & 0 \\
0 & 0 & 0 \\
0 & 0 & 0 \end{array}\right),
\label{eq6}
\end{equation}
\begin{equation}
\widetilde{N}_{\rm n}(r)=\left(\begin{array}{ccc}
n_{\rm n}(r)& 0 & 0 \\
0 & n_{\rm n}(r) & 0 \\
0 & 0 & n_{\rm n}(r) \end{array}\right),
\label{eq7}
\end{equation}
while $n_{\rm e}$ and $n_{\rm n}$ are the local electron and neutron
number densities, respectively.

\section*{THE PROCEDURE FOR CALCULATING THE FLUX OF STERILE NEUTRINOS}
Electron neutrinos from nonequilibrium neutronization
of matter are mostly emitted during the collapse
of an iron stellar core up to its stop and the
generation of a diverging shock front. In this case,
the fluxes of electron antineutrinos, along with neutrinos
of other flavors (muon and tau ones), may be
neglected at least for another several tens of milliseconds
after core bounce. Below we will describe
the procedure for calculating the generation of StN
fluxes under these conditions. We will distinguish two
cases: the case of high opacity with respect to electron
neutrinos and the transparent one. A spherical
symmetry is assumed with regard to the stellar core
geometry. The following neutrino--matter interaction
processes are taken into account in our calculations:
\begin{description}
\item[$\WID{\nu}{e}+n\rightleftarrows p+e^{-}$] \ the neutrino
absorption by a free neutron and the reverse process: the electron
capture by a proton with neutrino emission;
\item[$\WID{\nu}{e}+(A,Z)\rightleftarrows (A,Z{+}1)+e^{-}$]
\ the beta--processes on nuclei;
\item[$\WID{\nu}{e}+z\rightarrow\WID{\nu'}{e}+z'$]\ $z=n$ or
$p$ --- the elastic scattering by free nucleons;
\item[$\WID{\nu}{e}+(A,Z)\rightarrow\WID{\nu'}{e}+(A,Z)'$]
\ the coherent scattering by atomic nuclei. A very important source
of opacity due to the quadratic dependence of
the reaction cross section on the atomic weight
of the target:
$\WID{\sigma}{cs}\propto A^2$;
\item[$\WID{\nu}{e}+e^{-}\rightarrow\WID{\nu'}{e}+e'^{-}$]
\ the inelastic neutrino scattering by electrons. An important
process of thermalization of the neutrino radiation field.
\end{description}
The cross sections for these reactions were calculated
by Burrows and Thompson (2002) and Yudin and
Nadyozhin (2008) (for the neutrino scattering by an
electron).

\subsection*{THE CASE OF HIGH OPACITY}
When the density at the center of the stellar core
during its collapse reaches $\rho\sim 10^{13}$~\gccm, the neutrinos
turn out to be trapped. In this case, the neutrinos
propagate in the regime of diffusion, and the neutrino
mean free path $l_\nu$ in matter is small compared
to the characteristic length scales of the changes
in thermodynamic parameters (for more details, see
Imshennik and Nadyozhin 1972). The radius of this region in
the star is commonly called the neutrinosphere, by
analogy with the photosphere in ordinary stars. Let
us consider how StN are generated under these conditions.

We will proceed from the fact that we have a
function $\WID{P}{st}(\omega_\nu,\,x)$ giving the probability that an electron
neutrino with energy $\omega_\nu$, having traversed a distance
$x$, oscillated into sterile states (see Khruschov
et al. 2015). In this case, $\WID{P}{st}$ also depends on the local
thermodynamic parameters of matter such as the density, the
degree of neutronization, etc. which are assumed to
remain constant over the entire distance $x$. However,
the function  $\WID{P}{st}$ disregards the neutrino absorption
and scattering in matter, which ``knock out'' the neutrinos
from the beam. Consider this effect in detail.
Let we have a flux of electron neutrinos propagating
in a specified direction. The change in flux with
distance is described by the elementary solution of the
transfer equation $F_\nu(x)=F_\nu(0)e^{-\lambda_\nu x}$,
where  $\lambda_\nu\equiv 1/l_\nu$ is the absorptivity. In this case,
we disregard
the neutrino emission and scattering into a ``beam''
on the path length, because these neutrinos will be
incoherent with the original ones and will not be
involved in the oscillations.
For the same reason, the
opacity $\lambda_\nu(\omega_\nu)$ used here,
\begin{equation}
\lambda_\nu(\omega_\nu)=\WID{\lambda}{abs}(\omega_\nu)+\int\!\!
R^{\mathrm{out}}(\omega_\nu,\omega'_\nu,\eta)(1{-}f_\nu(\omega'_\nu))
d\omega'_\nu d\Omega',
\label{Lambda_def}
\end{equation}
differs from its ordinary value $\widetilde{\lambda}_\nu$ corrected
for the induced absorption (see Imshennik and Nadyozhin 1972)
by the factor $\widetilde{\lambda}_\nu=\lambda_\nu/(1{-}f_\nu^{\mathrm{eq}})$,
where $f_\nu^{\mathrm{eq}}$ is the
equilibrium neutrino distribution function at a given
point. In Eq. (\ref{Lambda_def}) the first and second terms describe
the neutrino absorption and scattering processes,
respectively, $R^{\mathrm{out}}$ is the scattering kernel dependent
on the energy of the initial neutrino $\omega_\nu$, the scattered
neutrino $\omega'_\nu$, and the cosine of the angle between
them $\eta$, while the integration is over the energy and
direction of the scattered neutrino (for more details,
see Yudin and Nadyozhin 2008). Note that the neutrino
distribution function $f_\nu(\omega'_\nu)$ is not necessarily
an equilibrium and isotropic one in the general case
and, in general, must be found from the solution of the
transport equation. However, under the conditions
we consider, inside the opaque region, the neutrino distribution is very close
to an equilibrium one.

Let us return to the consideration of the StN
generation. The fraction $\WID{P}{st}(x{+}dx){-}\WID{P}{st}(x)$ from the
electron neutrinos that traversed a distance $x$ oscillate
into sterile states in the range $(x\div x{+}dx)$. The total
StN flux generated by the flux of electron neutrinos propagating
in the medium is then
\begin{equation}
\WID{F}{st}=\int\limits_0^\infty F(x)\frac{d\WID{P}{st}(x)}{dx}dx=
F(0)\int\limits_0^\infty\WID{P}{st}(x)\lambda_\nu e^{-\lambda_\nu x}dx.
\label{F-G_function}
\end{equation}
As can be seen from (\ref{F-G_function}), we can introduce a function
$\WID{G}{st}(\omega_\nu)$ that defines the fraction of neutrinos from the
original beam that oscillated into sterile states:
\begin{equation}
\WID{G}{st}(\omega_\nu)\equiv\int\limits_0^\infty\WID{P}{st}(x,\omega_\nu)
e^{-x/l_\nu(\omega_\nu)}\dfrac{dx}{l_\nu(\omega_\nu)}\leq 1,
\label{G-function}
\end{equation}
where we used the identity $\lambda_\nu\equiv 1/l_\nu$ . The oscillation
length is seen to be effectively limited by a quantity of
the order of $l_\nu$. Naturally, apart from the neutrino energy,
the function $\WID{G}{st}$ also depends on the remaining
local parameters of matter. When deriving Eq. (\ref{G-function}),
we, first, assumed that $\WID{P}{st}(x)\ll 1$, a condition that
holds good in the case under consideration. Indeed,
for a reasonable choice of oscillation model parameters
the only region in the star where the oscillation
probability can be comparable to unity is the resonance
(in degree of neutronization $\tet$) zone (see Khruschov
et al. 2015). However, the electron neutrino
must have traversed a distance much greater than
the mean free path $l_\nu$ to oscillate into sterile states
with a high probability even in this case. Second, we
assumed that the parameters of matter did not change
significantly at distances of the order of the mean free
path. Thus, in particular, the condition for the density
gradient in the star
\begin{equation}
\left|\dfrac{d\ln\rho}{dr}\right|\ l_\nu\ll 1,
\label{eq11}
\end{equation}
and the analogous conditions for the remaining thermodynamic
quantities must hold. As our calculations
show, these conditions actually hold inside the neutrino
opacity region in the collapsing stellar core.

Let us now turn to the calculation of the neutrino
flux emitted from a unit volume of matter. This flux is
determined by the emissivity entering into the transport
equation (see Yudin and Nadyozhin 2008),
\begin{equation}
\kappa_\nu(\omega_\nu)=\left(1{-}f_\nu(\omega_\nu)\right)
\left[\lambda_\nu(\omega_\nu)e^{\frac{\mu_\nu{-}\omega_\nu}{kT}}
+\int\!\! R^{\mathrm{in}}(\omega_\nu,\omega'_\nu,\eta)
f'_\nu(\omega'_\nu)d\omega'_\nu d\Omega'\right]. \label{eq12}
\end{equation}
Here, the first and second terms in square brackets
are responsible for the neutrino emission and scattering
into a beam, respectively, $\mu_\nu$ is the neutrino
equilibrium chemical potential, and the factor $(1{-}f_\nu)$
describes the reduction of the phase space due to the
Pauli exclusion principle. It is easy to verify that the
Kirchhoff law holds: in equilibrium, when $f_\nu=f_\nu^{\mathrm{eq}}$,
\begin{equation}
\kappa_\nu=f_\nu^{\mathrm{eq}}\lambda_\nu.
\label{eq13}
\end{equation}
In particular, this is ensured by the symmetry properties
of the scattering kernel (Yudin and Nadyozhin
2008),
\begin{equation}
R^{\mathrm{in}}(\omega'_\nu,\omega_\nu,\eta)=R^{\mathrm{out}}
(\omega_\nu,\omega'_\nu,\eta),\quad R^{\mathrm{in}}
(\omega_\nu,\omega'_\nu,\eta)=e^{\frac{\omega'_\nu{-}\omega_\nu}{kT}}
R^{\mathrm{out}}(\omega_\nu,\omega'_\nu,\eta).
\label{eq14}
\end{equation}
Given the emissivity per unit volume of matter and
the fraction of neutrinos oscillating into sterile states
defined by the function $\WID{G}{st}(\omega_\nu,r)$ (\ref{G-function}),
we can find the StN energy flux at radius r in the star (see Ivanova
et al. 1969):
\begin{equation}
F^{\mathrm{st}}_\nu(r)=\int\limits_0^\infty\!\frac{\omega_\nu^3 d\omega_\nu}
{h^3c^2}\!\!\int\limits_0^r\!\!\kappa_\nu(\omega_\nu,r')\WID{G}{st}
(\omega_\nu,r')\dfrac{4\pi r'^2}{r^2} dr'. \label{F_st_star}
\end{equation}
The local StN number density can also be calculated:
\begin{equation}
n^{\mathrm{st}}_\nu(r)=\int\limits_0^\infty\!\frac{\omega_\nu^2
d\omega_\nu}{h^3c^3}\!\!\int\limits_{0}^{\WID{R}{s}}\!\!
\kappa_\nu(\omega_\nu,r')\WID{G}{st}(\omega_\nu,r')\frac{2\pi r'}{r}\ln\left|
\frac{r{+}r'}{r{-}r'}\right|dr'. \label{n_st_star}
\end{equation}
Note that the integral in (\ref{n_st_star}) is taken over the entire
star, while the integral in (\ref{F_st_star}) is taken only to the
current radius $r$. These expressions allow all the StN
flux characteristics of interest to us in the neutrino
opacity region to be found.

\subsection*{THE REGION OUTSIDE THE NEUTRINOSPHERE}
To calculate the StN generation in semitransparent
and transparent regions, i.e., where the electron
neutrino mean free path is comparable to or greater
than the characteristic length scales of the changes in
thermodynamic quantities, we first solve the transport
equation for electron neutrinos. The peculiarities
of the numerical scheme that we use can be found
in Nadyozhin and Otroshchenko (1980). Thus, we
determined the neutrino radiation field and, in particular,
the emissivity $\kappa_\nu$. However, in our case, apart
from the radial coordinate $r$ and energy $\omega_\nu$, the latter
also depends on the direction, i.e., on the direction
cosine $\eta$.

The same is also true for the function $\WID{G}{st}$ defining
the fraction of electron neutrinos oscillated into sterile
states: $\WID{G}{st}=\WID{G}{st}(\omega_\nu,r,\eta)$.
 Let us describe the algorithm
of finding it in this case. In each computational
domain with coordinate $r$ we determine the set of
directions, i.e., the direction cosines $\eta_i$, $(i=1\div 6)$,
chosen as the grid points of Legendre polynomials
for the subsequent numerical integration using a 6-point scheme.
The equation for the propagation of
a beam of electron neutrinos with a given energy $\omega_\nu$
through the star is numerically solved along each of
the directions. Initially, the normalized set of fluxes of
all neutrino flavors is
$\vec{F}_\nu\equiv\left\{\WID{F}{e},\WID{F}{\mu},\WID{F}{\tau},\WID{F}{x},
\WID{F}{y},\WID{F}{z}\right\}=\left\{1,0,0,0,0,0\right\}$. In each
computational interval in the star the flux of electron neutrinos decreases
due to their absorption and scattering:
$\WID{F^{{\kern1pt}\prime}}{e}=\WID{F}{e}e^{{-}\bigtriangleup l/l_\nu}$
where $\bigtriangleup l$ is the length of the interval along the propagation
path (recall that we disregard the reverse
processes, i.e., the emission and scattering into a
beam, because these neutrinos will be incoherent
with the original ones). Thereafter, we determine
the redistribution of fluxes between the neutrino flavors
using the function $\WID{P}{st}(\omega_\nu,\bigtriangleup l)$. By considering
$\WID{P}{st}$ as an operator,
we can symbolically write $\vec{F}^{{\kern1pt}\prime}_\nu=
\WID{\widehat{P}}{st}(\omega_\nu,\bigtriangleup l)\vec{F}_\nu$.
The two processes described above,
namely the knocking-out of neutrinos from the beam
and the oscillation redistribution of flavors, allow the
local value of the function $\WID{G}{st}$ to be found. The
computation is continued until the fraction of electron
neutrinos in the beam drops below a certain level (we
took $5\%$ of the sum of the fluxes of all neutrino flavors)
or until the stellar surface is reached.

\begin{figure}[htb]
\epsfxsize=0.5\textwidth
\hspace{-2cm}\center\epsffile{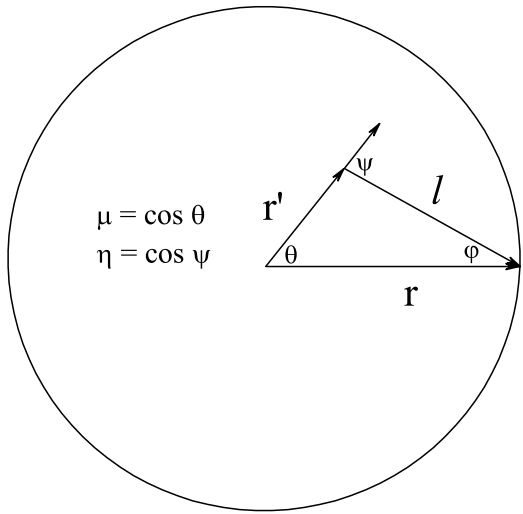}
\caption{\rm Scheme to calculate the flux of sterile neutrinos.}
\label{Fig.Scheme}
\end{figure}
We now have all information to calculate the StN
energy flux at a distance $r$ from the stellar center.
The previous expression (\ref{F_st_star}) must be replaced by the
following integral (see Fig.~\ref{Fig.Scheme} for an explanation):
\begin{equation}
F^{\mathrm{st}}_\nu(r)=\int\limits_0^\infty\!\frac{\omega_\nu^3 d\omega_\nu}
{h^3c^2}\!\!\int\limits_0^{\WID{R}{s}}\!\!\int\limits_{-1}^1\!\!
\Upsilon^{\mathrm{st}}_\nu(\omega_\nu,r',\eta)
\dfrac{2\pi r'^{{\kern1pt}2}\cos\varphi}{l^2}dr'd\mu, \label{F_st_star_out}
\end{equation}
where we, for brevity, denoted
$\Upsilon^{\mathrm{st}}_\nu\equiv\kappa_\nu\WID{G}{st}$.
Here, $\mu=\cos\theta$,
$l^2=r^2{+}r'^{{\kern1pt}2}{-}2rr'\mu$, $\cos\varphi=(r{-}r'\mu)/l$, and
$\eta=\cos\psi=(r\mu{-}r')/l$. Remarkably,
the integration over the angles in (\ref{F_st_star_out}) can be
performed explicitly. Indeed, it is easy to show that
\begin{equation}
d\eta=d\left(\dfrac{r\mu-r'}{l}\right)=\dfrac{r^2\cos\varphi}{l^2}d\mu.
\label{eq18}
\end{equation}
As $\mu$ changes within the range $(-1, 1)$ $\eta$ changes
from $-1$ to $(r{-}r')/|r{-}r'|$. This means that at $r'>r$
the integral over the angle in (\ref{F_st_star_out})
is zero. Finally, we can write
\begin{equation}
F^{\mathrm{st}}_\nu(r)=\int\limits_0^\infty\!\frac{\omega_\nu^3 d\omega_\nu}
{h^3c^2}\!\!\int\limits_0^r\!\!\Upsilon^{\mathrm{st}}_{\nu 0}(\omega_\nu,r')
\dfrac{4\pi r'^{{\kern1pt}2}}{r^2}dr', \label{F_st_star_out_fin}
\end{equation}
where we introduced the following notation for the
zeroth angular moment of
$\Upsilon^{\mathrm{st}}_{\nu}$:
\begin{equation}
\Upsilon^{\mathrm{st}}_{\nu 0}(\omega_\nu,r')\equiv\dfrac{1}{2}
\int\limits_{-1}^{1}\Upsilon^{\mathrm{st}}_{\nu 0}(\omega_\nu,r',\eta)d\eta.
\label{Upsilon_0}
\end{equation}
It is important to note that, just as (\ref{F_st_star}), our final
expression for the flux (\ref{F_st_star_out_fin}) includes the integration
only to the current $r$ and not to the full stellar radius
$\WID{R}{s}$. In addition,
a remarkable feature of Eq.~(\ref{F_st_star_out_fin})
is that the quantity $\Upsilon^{\mathrm{st}}_{\nu}$, which,
recall, is equal to the product of the emissivity of matter by the StN
fraction, enters only by its minimal (zeroth) moment
of the distribution. We perform the numerical angular
integration in (\ref{Upsilon_0}) using a six-point Gaussian
scheme.

\subsection*{The $\WID{\nu}{e}{-}\WID{\nu}{e}$ SCATTERING EFFECT}
So far we have taken into account only the neutrino
scattering by the following components of matter
when calculating the neutrino oscillation effects
in the medium: neutrons, protons, and electrons.
This led to the dependence of the interaction potential
$\WID{V}{s} \propto 3\WID{Y}{e}{-}1$ (see, e.g., Abazajian et al. 2012), where
$Y_{\rm e}$ is the ratio of the electron number density to the
total number density of neutrons and protons. Given
that $\WID{Y}{e}=\frac{1}{1+\tet}$, as would be expected, we obtain the
critical value $\WID{V}{s}=0$ at $\tet=2$. However, if there is
a considerable number of electron neutrinos in the
medium (as in the opaque region of the central supernova
core), then the possibility of neutrino--neutrino
scattering should also be taken into account. This
leads to a modification of the potential (see Blinnikov
and Okun 1988): $\WID{V}{s}\propto 3\WID{Y}{e}{-}1{+}4Y_\nu$, where
$Y_\nu$ is the dimensionless number density of electron neutrinos
in the medium (see Eq.~(\ref{Y_nu}) below). Naturally, the
oscillation function $\WID{G}{st}$ begins to also depend on $Y_\nu$ in
this case. The critical value of  $\tet$ is shifted to values
greater than two:
\begin{equation}
\WID{\theta}{cr}=2+\dfrac{12Y_\nu}{1-4Y_\nu}. \label{eta_Ynu}
\end{equation}
However, an important clarification should be made
here: Eq.~(\ref{eta_Ynu}) and the above formula for the potential
$\WID{V}{s}$ containing $Y_\nu$ are valid only for an isotropic distribution
of neutrinos. In our case, however, the angular
neutrino distribution function can deviate greatly
from an isotropic one. In particular, outside the
neutrinosphere the neutrinos mostly propagate in a
narrow solid angle in a direction away from the stellar
center. Let us show how the expression for the critical
degree of neutronization $\WID{\theta}{cr}$ should be modified in this
case.

To begin with, let us introduce the angular moments
of the neutrino distribution function according
to the definition
\begin{equation}
f_{\nu k}=\dfrac{1}{2}\int\limits_{-1}^{1}\mu^{k}f_\nu(\mu)d\mu,
\label{eq22}
\end{equation}
where $\mu$ is the cosine of the angle between the neutrino
propagation direction and the radius vector of a
given point in the star. The quantity $Y_\nu$ is expressed
via the zeroth moment of the distribution function as
\begin{equation}
Y_\nu=\dfrac{4\pi\WID{m}{u}}{\rho(h c)^3}\!\!
\int\limits_0^\infty\!\omega_\nu^2 f_{\nu 0}(\omega_\nu)d\omega_\nu,
\label{Y_nu}
\end{equation}
where $\WID{m}{u}$ is an atomic mass unit.

\begin{figure}[htb]
\epsfxsize=0.4\textwidth
\hspace{-2cm}\center\epsffile{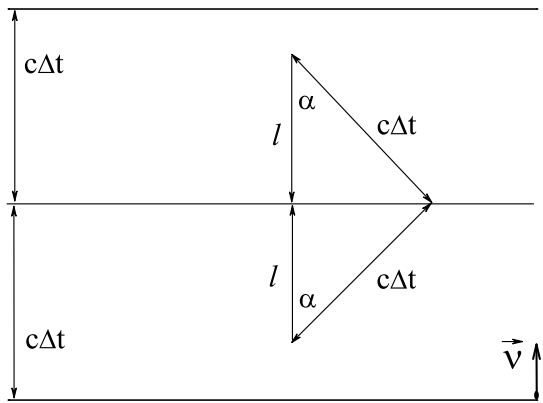}
\caption{\rm Illustration to the calculation of $\WID{\theta}{cr}$
when the $\WID{\nu}{e}{-}\WID{\nu}{e}$ scattering is taken into account.}
\label{Fig.cdt}
\end{figure}
Let us now consider a neutrino propagating with
speed $c$ for a time $\triangle t$ ``upward'' in Fig.~\ref{Fig.cdt}. It can
interact with all the neutrinos from the lower halfplane
moving ``downward''. In addition, in the lower
half--plane it can interact with the upward--moving
neutrinos that were at a distance $l$ from the middle
line at $t = 0$ and whose propagation direction made
an angle no smaller than $\alpha$ with the direction of the
neutrino under consideration. Similarly, from the
upper half--plane our neutrino can interact only with
the neutrinos that propagate downward within the
angle  $\alpha$ (see Fig.~\ref{Fig.cdt}). Thus, the quantity
$4\pi f_{\nu 0}$ in (\ref{Y_nu}) must be replaced by
\begin{equation}
\int\limits_{\Omega_1}\!f_\nu d\Omega+\frac{1}{c\triangle t}\!
\int\limits_0^{c\triangle t}\!\Bigg(\,\int\limits_{\Omega_2}\!f_\nu d\Omega
+\int\limits_{\Omega_3}\!f_\nu d\Omega\Bigg)dl, \label{fdOmega}
\end{equation}
where the integration over the solid angle $\Omega$ is within
the following limits:
\begin{equation}
\begin{aligned}
&\Omega_1:\quad (\vec{n}\vec{m})\leq 0,\\
&\Omega_2:\quad 0\leq(\vec{n}\vec{m})\leq
\frac{l}{c\triangle t},\\
&\Omega_3:\quad -1\leq(\vec{n}\vec{m})\leq -\frac{l}{c\triangle t}.
\end{aligned}
\label{eq25}
\end{equation}
Here $\vec{n}$ and $\vec{m}$
are unit vectors in the propagation
directions of the original neutrino and the neutrino
of the medium, respectively. Let us decompose the
neutrino distribution function $f_\nu$ into its angular moments:
$f_\nu=f_{\nu 0}+3\mu f_{\nu 1}+O(\mu^2)$. The cosine of
the angle $\mu$ can be expressed via the angle between
the interacting neutrinos $\chi\equiv(\vec{n}\vec{m})$ and the direction
cosine of the original neutrino $\eta$ as
\begin{equation}
\mu=\chi\eta+\sqrt{1{-}\chi^2}\sqrt{1{-}\eta^2}\cos(\varphi_1{-}\varphi_2).
\label{eq26}
\end{equation}
This allows the integration in (\ref{fdOmega}) to be performed analytically
using the relation $d\Omega=d\varphi_2 d\chi$. The critical
degree of neutronization now depends on the propagation
direction of the neutrino under consideration,
i.e., this is the function $\WID{\theta}{cr}(\eta)$, and it is still
defined by Eq.~(\ref{eta_Ynu}), but with $Y_\nu$
replaced by $\widehat{Y}_\nu(\eta)$, where
\begin{equation}
\widehat{Y}_\nu(\eta)=\dfrac{4\pi\WID{m}{u}}{\rho(h c)^3}\!\!
\int\limits_0^\infty\!
\big(f_{\nu 0}(\omega_\nu){-}\eta f_{\nu 1}(\omega_\nu)\big)\omega_\nu^2
d\omega_\nu.\label{Y_nu_2}
\end{equation}
Expression (\ref{Y_nu_2}) is valid to within $O(f_{\nu 3})$, because
the contribution of the second moment $f_{\nu 2}$ can be
shown to be zero. In the opaque region $f_{\nu 1}\ll f_{\nu 0}$ and
(\ref{Y_nu_2}) is reduced to (\ref{Y_nu}). In contrast, in the region
high above the neutrinosphere $f_{\nu 1}\approx f_{\nu 0}$
and $\widehat{Y}_\nu(\eta)$ is highly anisotropic:
if $\eta\approx 1$ then $\widehat{Y}_\nu(\eta)\approx 0$,
because all neutrinos (including that under consideration)
move nearly radially away from the stellar center
without any interaction. On the other hand, at $\eta\approx -1$ the
neutrino under consideration moves toward the flux
of remaining neutrinos, and the interaction effect is
doubled.

As we will show below, the $\WID{\nu}{e}{-}\WID{\nu}{e}$ scattering effect
itself turns out to be very important in the calculations
of the StN fluxes from supernovae. It can lead to both
an increase and a decrease in the StN yield. Below we
will distinguish three cases: the calculations without
any scattering effect $(\WID{\theta}{cr}=2)$, with isotropic scattering
($\WID{\theta}{cr}$ is given by Eq.~(\ref{eta_Ynu}) with $Y_\nu$ determined
from Eq.~(\ref{Y_nu})), and with anisotropic scattering with
Eq.~(\ref{Y_nu_2}) for $Y_\nu\equiv \widehat{Y}_\nu$ in Eq.~(\ref{eta_Ynu})
for $\WID{\theta}{cr}$.

\section*{RESULTS}
We used the results of hydrodynamic simulations
for the collapse of an iron stellar core with a mass
of $2M_\odot$ from Yudin (2009) as a basis for our calculations
of the StN fluxes from a supernova. From
the above paper we took the distributions of thermodynamic
and hydrodynamic parameters of the core
at various instants of time, starting from the instant
the stellar core lost its stability up to several tens of
milliseconds after bounce. This time interval corresponds
to the formation of the first powerful peak of
neutrino emission consisting mostly of electron neutrinos
from nonequilibrium neutronization of matter.
In this case, the degree of neutronization $\tet$ increases
from $\tet\approx 1$ in the pre--collapse matter to $\tet\approx 2.5$ in
the central core region (see also Liebend\"{o}rfer 2005),
while in the neutronization region behind the expanding
shock front (see Figs ~\ref{Fig.1500_profile}
and \ref{Fig.35000_profile} below) $\tet$ reaches
even greater values. The collapsing matter, naturally,
passes through the most interesting region of the
critical degree of neutronization, i.e., the zone of a
resonant enhancement of the neutrino oscillations.

\subsection*{THE PREBOUNCE COLLAPSE PHASE}
\begin{figure}[htb]
\epsfxsize=1.\textwidth %17cm
\hspace{-2cm}\center\epsffile{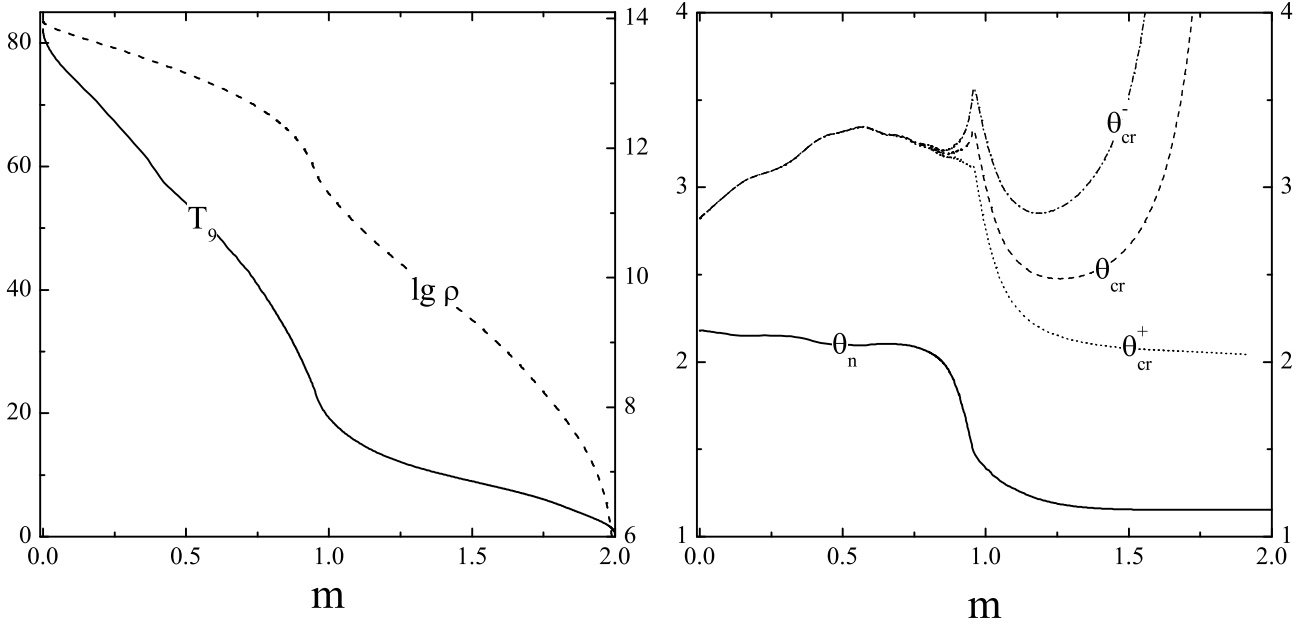}
\caption{\rm (a: left) Temperature (solid curve, left axis) and logarithm
of the density (dashed curve, right axis) versus mass coordinate
$m=M/M_\odot$ inside the collapsing stellar core. (b: right)
Degree of neutronization $\tet$ and its critical values with allowance
for the $\WID{\nu}{e}{-}\WID{\nu}{e}$ scattering.}
\label{Fig.1500_profile}
\end{figure}

Let us first consider the stellar core collapse phase
before bounce and the formation of a diverging shock
front. The distribution of parameters in the stellar
core at the moment when the density of matter at the
center is lower than the nuclear density
$\WID{\rho}{n}\approx 2.6\times 10^{14}$~\gccm\
by only several times is shown in Fig.~\ref{Fig.1500_profile}.
Figure~\ref{Fig.1500_profile}a shows the temperature
$T_9\equiv T\times 10^{-9}$~K (solid curve, left axis)
and the logarithm of the density
$\lg\rho$ (dashed curve, right axis) as functions of the
mass coordinate $m=M/M_\odot$. At $m\approx 0.85$ we can
clearly see the formation of a density jump, the place
where the future shock front is formed. Approximately
the same mass coordinate separates the central neutronized
part of the star with $\tet>2$ from the outer
region with $\tet\approx 1$ ($\tet$ is indicated by the solid curve
in Fig.~\ref{Fig.1500_profile}b). The critical degrees of neutronization
with allowance made for the $\WID{\nu}{e}{-}\WID{\nu}{e}$ scattering are
also presented here. $\WID{\theta}{cr}$ corresponds to the critical
value for isotropic scattering, while $\WID{\theta}{cr}^{\pm}$ corresponds to
anisotropic scattering for $\eta=\pm 1$ in Eq.~(\ref{Y_nu_2}). In the
outer part of the core $\WID{\theta}{cr}$ and $\WID{\theta}{cr}^{-}$
increase rapidly, because $Y_{\nu}$ increases. Indeed,
$Y_\nu\equiv \dfrac{\WID{m}{u} n_\nu}{\rho}$, where $n_\nu$ is
the neutrino number density that drops with distance
in the outer part approximately as $n_\nu\sim 1/r^2$. Since
the density decreases more rapidly, $Y_\nu$ increases. This
effect is compensated only for  $\WID{\theta}{cr}^{+}$, because in the outer
region $f_{\nu 0}\approx f_{\nu 1}$ (see Eq.~(\ref{Y_nu_2})).

\begin{figure}[htb]
\epsfxsize=1.\textwidth
\hspace{-2cm}\center\epsffile{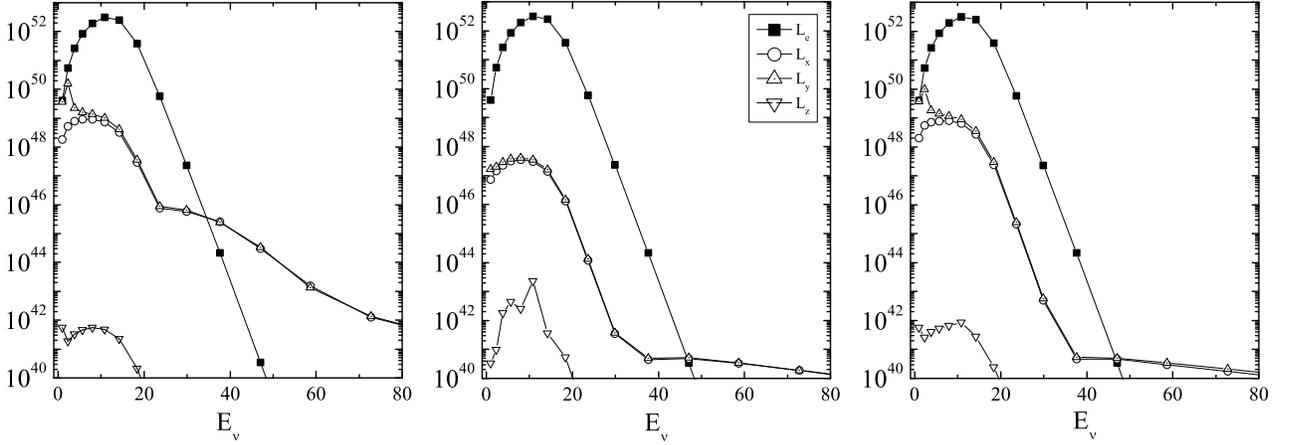}
\caption{\rm Spectral luminosities  $L_\nu$ (erg s$^{-1}$ MeV$^{-1}$) of
sterile neutrinos leaving the collapsing stellar core.
The calculations (a: left) without $\WID{\nu}{e}{-}\WID{\nu}{e}$ scattering,
(b: center) with isotropic scattering, and (c: right)
with anisotropic scattering.}
\label{Fig.1500_spectrum}
\end{figure}
Let us now consider the results of our calculations
of the StN generation with this stellar profile.
Figure~\ref{Fig.1500_spectrum}
shows the spectral luminosities $L_\nu$ (erg s$^{-1}$ MeV$^{-1}$)
of neutrinos leaving the collapsing core. The curve
with filled squares indicates the spectrum of electron
neutrinos; the remaining curves indicate the spectra of
the $x$, $y$, and $z$ StN components.
Figures \ref{Fig.1500_spectrum}a--\ref{Fig.1500_spectrum}c correspond
to the calculations without $\WID{\nu}{e}{-}\WID{\nu}{e}$ scattering,
with isotropic scattering, and with anisotropic
scattering, respectively. The spectra of the $x$ and
$y$ components virtually coincide in the entire range
of energies, except for the lowest ones $(E_\nu\lesssim 5$~MeV),
while the yield of the $z$ component is much smaller.
The StN spectrum itself has a characteristic shape:
it consists of a central part with a maximum at $
\sim 10$~MeV and a broad ``pedestal'' with a considerably
flatter spectrum at high energies. This part of the
spectrum is attributable to the StN produced in the
deepest and hottest regions of the stellar core. On
the whole, the StN yield is smaller than the yield of
the active component (electron neutrinos) by more
than two orders of magnitude. At high energies
($E_\nu\gtrsim 40$~MeV) the spectral luminosity of the sterile
component begins to dominate.

Let us now consider the scattering effect. Comparison
of Figs.~\ref{Fig.1500_spectrum}a and ~\ref{Fig.1500_spectrum}b
shows that the absence of a resonance reduces considerably the StN yield.
Indeed, $\WID{\theta}{cr}$ when the $\WID{\nu}{e}{-}\WID{\nu}{e}$
scattering is taken into account is too large compared to the typical
$\WID{\theta}{cr}=2$ (see Fig.~\ref{Fig.1500_profile}b). Allowance for
the anisotropy (Fig.~\ref{Fig.1500_spectrum}c) gives
a mixed spectrum: the range of low energies is analogous
to the case without scattering; the high--energy
spectrum (the so--called ``pedestal'') is suppressed, as
in the isotropic case. This is explained by a combination
of two effects: first, different generation depths
of neutrinos with different energies (high--energy StN
are generated in deep layers of the core) and, second,
the dependence of the resonant enhancement of oscillations
on the neutrino energy.
\begin{figure}[htb]
\epsfxsize=0.6\textwidth
\hspace{-2cm}\center\epsffile{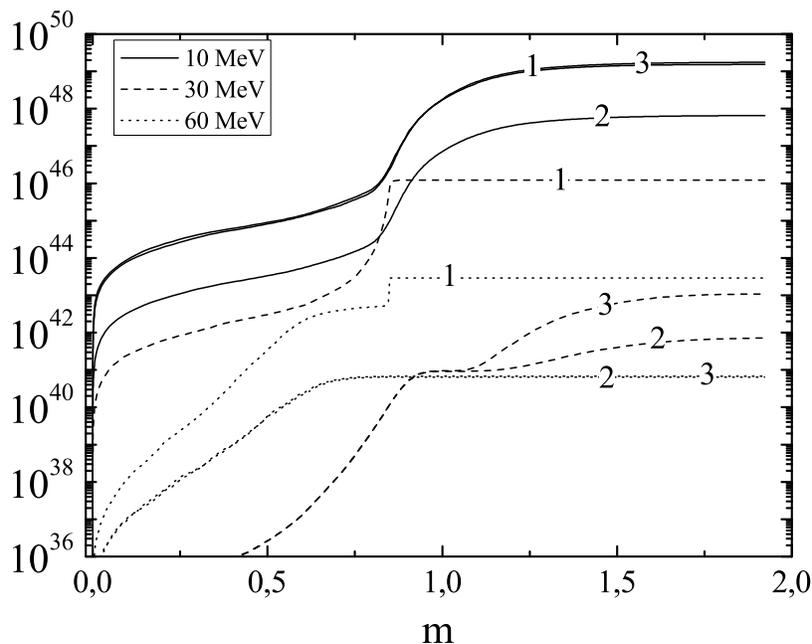}
\caption{\rm Total local luminosities of sterile neutrinos $L_\nu(m)$
versus mass coordinate $m=M/M_\odot$ in the collapsing stellar core
for $E_\nu=10, 30$ and $60$~MeV (the solid, dashed, and dotted curves,
respectively). The numbers indicate the cases without scattering (1),
with isotropic scattering (2), and with anisotropic scattering (3).}
\label{Fig.1500_Fluxes}
\end{figure}
To demonstrate this,
we provide Fig.~\ref{Fig.1500_Fluxes}, which shows the total spectral StN
luminosities $L_\nu=\WID{L}{x}+\WID{L}{y}+\WID{L}{z}$ as functions of the
mass coordinate $m=M/M_\odot$ in the collapsing stellar
core. The spectral luminosity at a given stellar radius
$r$ is defined via the local spectral flux by the formula
$L_\nu=4\pi r^2 F_\nu(r)$ (erg s$^{-1}$ MeV$^{-1}$). The luminosities
are shown for $E_\nu=10, 30$ and $60$~MeV (the solid,
dashed, and dotted curves, respectively). The numbers
mark the cases without scattering (1), with isotropic
scattering (2), and with anisotropic scattering (3).
The resonance region is easily seen on curves 1, but
it is smeared at low energies. At an energy of $30$~MeV
(dashed lines) the resonance zone is still fairly wide
and leads to an increase in the luminosity by almost
two orders of magnitude (there is nothing of the kind
on dashed curves 2 and 3, i.e., with scattering). At
$E_\nu=60$~MeV without scattering the resonance is
narrow and leads to an increase in the luminosity
approximately by a factor of 4, while the curves with
scattering (2 and 3) exhibit a gradual rise without
resonances.

Thus, the following conclusions can be reached:
first, the $\WID{\nu}{e}{-}\WID{\nu}{e}$ scattering effect is very
important in the calculations of the StN fluxes, because it changes
the resonant degree of neutronization. In the case
under consideration, the scattering has a negative effect
on the generation of sterile neutrino components,
but, as we will see below, this is not always the case.
Second, the resonant enhancement of oscillations at
the critical degree of neutronization pointed out in
Khruschov et al. (2015) actually plays an important
role under stellar core collapse conditions. Third, this
effect depends strongly on the neutrino energy: for
very high energies the resonance zone is too narrow
(because the mean free path is small), while
for low energies the effect itself is small. An optimum
is probably reached in the energy range $E_\nu=(20\div40)$~MeV, which
determines the region where a
change in the pattern of the spectrum in Fig.~\ref{Fig.1500_spectrum}, i.e.,
a transition to a more gradual drop in luminosity with
energy (the ``pedestal'' effect), occurs.

\subsection*{THE POSTBOUNCE COLLAPSE PHASE}
\noindent
Let us now consider the supernova explosion
phase that begins after an abrupt stop of the core
collapse, bounce, and the generation of a diverging
shock front that, having reached the outer stellar
layers, must eject them, producing the observed
explosion. The passage of the shock front through
the matter heats it up, breaking all complex nuclei
into free neutrons, protons, and $\alpha$--particles. Abrupt
neutronization of matter leading to $\tet\sim 10$ also
occurs in the postshock region. Consider this phase
in detail.

\begin{figure}[htb]
\epsfxsize=1.\textwidth
\hspace{-2cm}\center\epsffile{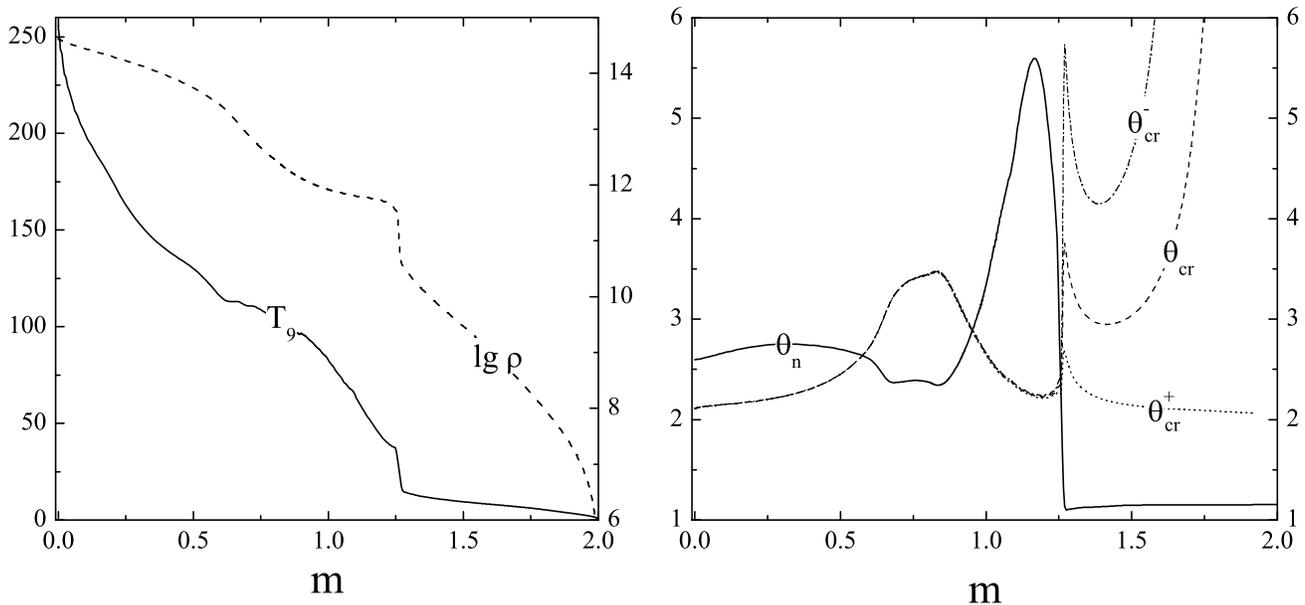}
\caption{\rm Parameters in the core about 3 ms after bounce.
(a: left) temperature (solid curve, left axis) and logarithm of the density
(dashed curve, right axis) versus mass coordinate $m=M/M_\odot$ inside
the collapsing stellar core. (b: right) degree of neutronization
$\tet$ and its critical values with allowance made
for the $\WID{\nu}{e}{-}\WID{\nu}{e}$ scattering.}
\label{Fig.35000_profile}
\end{figure}
\begin{figure}[htb]
\epsfxsize=0.6\textwidth
\hspace{-2cm}\center\epsffile{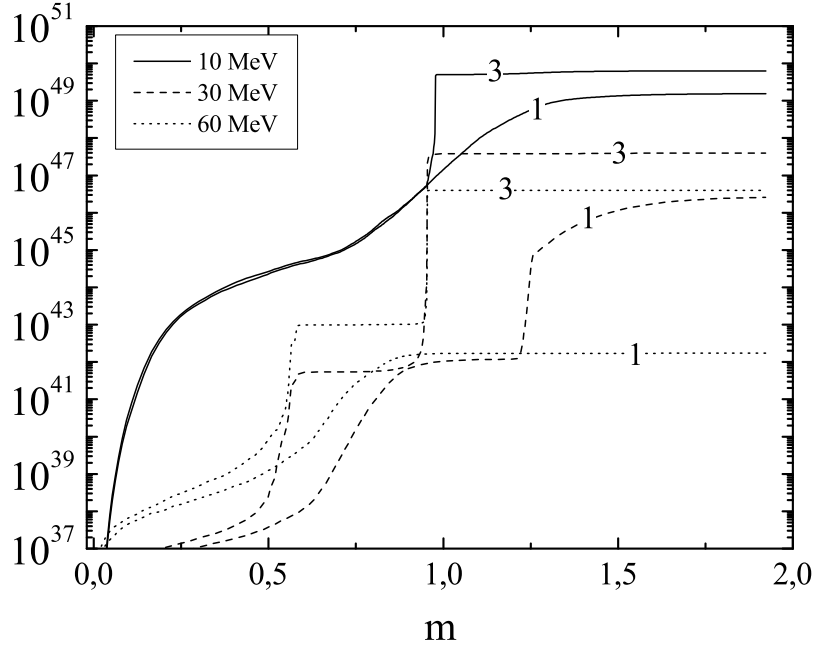}
\caption{\rm Total local luminosities of sterile neutrinos $L_\nu(m)$
versus mass coordinate $m=M/M_\odot$ in the collapsing stellar core
for $E_\nu=10, 30$ and $60$~MeV (the solid, dashed, and dotted curves,
respectively). The numbers indicate the cases without
scattering (1) and with anisotropic scattering (3).}
\label{Fig.35000_Fluxes}
\end{figure}

In Fig.~\ref{Fig.35000_profile}, which is analogous
to Fig.~\ref{Fig.1500_profile}, the parameters
of the stellar core are shown at approximately
3 ms after bounce. The shock front seen from the
jumps in all thermodynamic quantities is at $m \approx 1.25$.
The matter in the region traversed by the shock front,
i.e., in the range $0.8\leq m\leq 1.25$, is strongly neutronized
(see Fig.\ref{Fig.35000_profile}b). The situation in the core after its
bounce is seen to have changed significantly. Without
$\WID{\nu}{e}{-}\WID{\nu}{e}$ scattering the curve $\tet(m)$ still intersects the
resonant value $\WID{\theta}{cr}=2$ only once at the shock front.
However, with scattering $\WID{\theta}{cr}(m)$ now intersects with
the curve $\tet(m)$ three times! The result can be seen in
Fig.~\ref{Fig.35000_Fluxes}, which is analogous to Fig.~\ref{Fig.1500_Fluxes}
for the prebounce collapse phase, except that only two cases are presented
here: without scattering and with anisotropic scattering.
We will begin from $E_\nu=10$~MeV (solid
lines). The total spectral luminosities for cases 1 and
3 coincide up to the zone of the second resonance at
$m\approx 0.95$, whereupon the luminosity with scattering
increases by almost three orders of magnitude. The
first resonance at $m\approx 0.57$ and the third one at the
shock front do not give any noticeable enhancement
of oscillations. At an energy of $30$~MeV (dashed curves)
for the case without scattering the resonance at $\theta=2$
manifests itself at the shock. The first and the second
resonances for case 3 are also clearly seen. They
also work for an energy of $60$~MeV (dotted curves),
while for case 1 at this energy the resonance at the
shock vanishes. All of this draws a complex picture
of the resonant enhancement of oscillations whose
properties depend on the stellar core parameters, the
neutrino energy under consideration, and the included
oscillation parameters.

\begin{figure}[h!tb]
\epsfxsize=0.6\textwidth
\hspace{-3cm}\center\epsffile{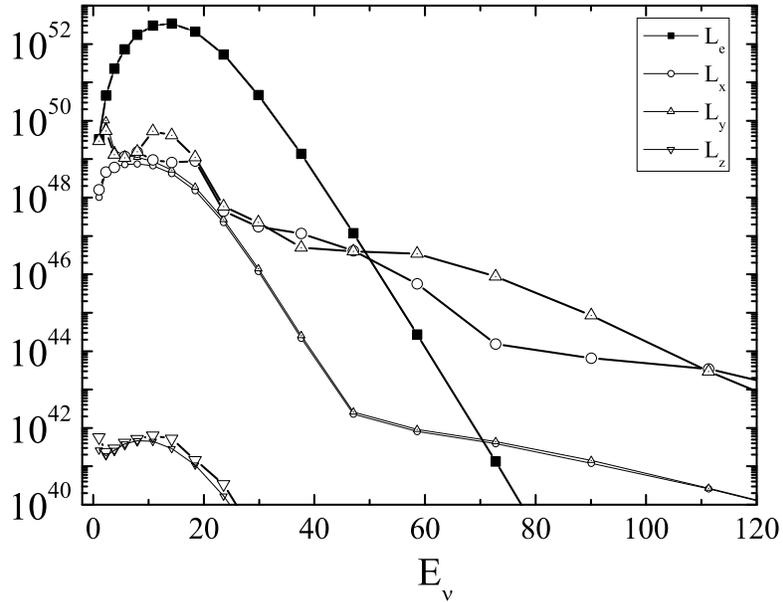}
\caption{\rm Spectral luminosities  $L_\nu$ (erg s$^{-1}$ MeV$^{-1}$)
of sterile neutrinos leaving the collapsing stellar core. The black filled
squares indicate the spectrum of electron neutrinos. The small symbols
and thin curves represent the calculation without
$\WID{\nu}{e}{-}\WID{\nu}{e}$ scattering; the large symbols
and thick curves correspond to anisotropic scattering.}
\label{Fig.35000_Spectra}
\end{figure}
The spectral characteristics of the fluxes of neutrinos
leaving the stellar core at the instant of time under
consideration are shown in Fig.~\ref{Fig.35000_Spectra}.
Just as in Fig.~\ref{Fig.1500_spectrum},
the curve with filled squares indicates the spectrum of
electron neutrinos. The small symbols and thin curves
indicate the result of our calculation without $\WID{\nu}{e}{-}\WID{\nu}{e}$
scattering; the large symbols and thick curves represent
our calculation with anisotropic scattering. In
principle, the StN spectrum without $\WID{\nu}{e}{-}\WID{\nu}{e}$ scattering
is very similar to the results of our calculation with
the profile before bounce discussed above, while the
spectrum with anisotropic scattering differs significantly.
First, it exhibits some oscillations between
the $x$ and $y$ StN components (their sum demonstrates a
much smoother behavior). Second, the spectrum
itself is considerably harder, and the ``pedestal'' effect
is not such pronounced. This is apparently explained
by the presence of several resonances lying at different
depths in the stellar core. The deeper the
resonance region lies, the more energetic StN can be
produced there in significant quantity, with all the
reservations concerning the dependence of the oscillation
strength on the neutrino energy made above.
Despite a considerable complication of the picture
of oscillations, the situation qualitatively remains as
before: the maximum of the spectrum occurs at low
energies $E_\nu<20$~MeV, with the total luminosity of
sterile components being lower than the luminosity
of electron neutrinos by several orders of magnitude.
However, at high energies ($E_\nu\gtrsim 50$~MeV) the StN
spectrum falls off much more gently, and the StN
luminosity exceeds the luminosity of the active component.

\subsection*{CALCULATIONS WITH OTHER PARAMETERS OF STERILE NEUTRINOS}
The StN parameters that we used in our calculations
as the ``main'' ones are not the only possible
ones. It is very interesting to find out the possible
influence of other sets of parameters (mixing
angles, masses, etc.) on the properties of the resulting
neutrino emission from supernovae. We will
consider the case of inverted hierarchy (IH) of the
neutrino mass spectrum as maximally differing from
the ``main'' one. The question about the hierarchy of
the active neutrino mass spectrum has not yet been
solved; therefore, the IH case should be considered
as an admissible one. We used the parameters for
this case from Zysina et al. (2014) and Khruschov
et al. (2015).

\begin{figure}[htb]
\epsfxsize=0.6\textwidth
\hspace{-3cm}\center\epsffile{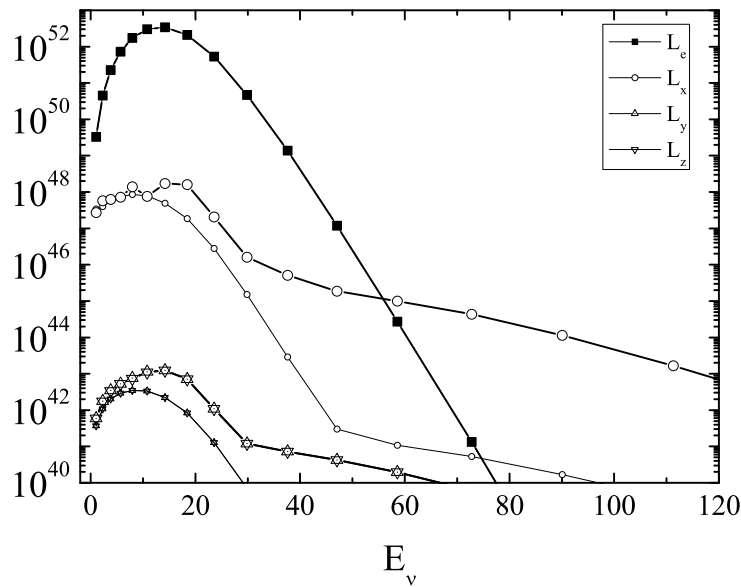}
\caption{\rm Spectral luminosities $L_\nu$ (erg s$^{-1}$ MeV$^{-1}$)
of sterile neutrinos leaving the collapsing stellar core for the case
of inverted hierarchy of the active neutrino masses. The black filled
squares indicate the spectrum of electron neutrinos. The small symbols
and thin lines represent the calculation without
$\WID{\nu}{e}{-}\WID{\nu}{e}$ scattering; the large symbols and
thick lines correspond to anisotropic
scattering.}
\label{Fig.35000_Spectra_IH}
\end{figure}
To demonstrate the influence of this set of parameters
on the StN generation in a supernova, we will
present the results of our calculations under the same
supernova core conditions 3 ms after bounce as those
in Fig.~\ref{Fig.35000_profile}.
Figure~\ref{Fig.35000_Spectra_IH} shows the StN spectra for the
IH case, with all designations being the same as those
in Fig.~\ref{Fig.35000_Spectra}. As can be seen, the picture is generally
preserved. However, the yields of $y$-- and $z$--type StN
with almost the same value are now suppressed. This
set of parameters leads to a reduction in StN generation
by more than an order of magnitude, although the
spectra are qualitatively very similar to the results of
our calculations for the ``main'' NH case (see Fig.~\ref{Fig.35000_Spectra}).

Apart from the consideration of other mixing
parameters within the phenomenological oscillation
model being investigated, a modification of the model
itself is also possible, for example, by including the
new hypothetical interaction between sterile neutrinos
(see, e.g., Abazajian et al. 2012; Chu et al. 2015).
This effect is capable of affecting the quantitative
oscillation characteristics, but its study is beyond the
scope of this paper.

\section*{CONCLUSIONS}
In this paper we developed and implemented a
combined algorithm for calculating the generation of
sterile neutrinos inside the collapsing supernova core.
It includes both a direct calculation of the neutrino
oscillations and allowance for the generation, propagation,
and interaction of neutrinos in a medium. This
algorithm includes a separate consideration of both
the region opaque to neutrinos (under neutrinosphere) and
the semitransparent shell surrounding it. Basically,
it is an efficient simplified method of calculating the
oscillations in a medium that is applicable when the
resulting yield of sterile neutrinos is small. Otherwise,
the complete system of kinetic equations for all
neutrino flavors (see, e.g., Rudzsky 1990) should be
solved.

Admittedly, the main result of our calculations is
the demonstration that the derived energy characteristics
of the StN flux are relatively small: for example
the total StN luminosity turns out to be at least several
orders of magnitude lower than the luminosity of
electron neutrinos. In this way, we not only confirmed
the applicability of the algorithm under consideration
but also showed that sterile neutrinos, in contrast to
the assumption made by Hidaka and Fuller (2006,
2007), are incapable of affecting significantly the collapse.
At least, this is true for the parameters and
the oscillation model we considered.

An important result of our calculations is the proof
that the resonant enhancement of StN oscillations
noted previously in Khruschov et al. (2015) actually
occurs inside the collapsing star. Each such
resonance is capable of increasing the luminosity by
several orders of magnitude. We also showed that
the $\WID{\nu}{e}{-}\WID{\nu}{e}$ scattering
effect could affect significantly
the radiation parameters of the sterile component.
In particular, at the postbounce phase it leads to an
increase in the number of resonances in degree of
neutronization.

Besides, we showed that the choice of other StN
model parameters (hierarchy, mixing angles, neutrino
masses) could affect significantly the parameters of
the StN flux. We are going to systematically study
the domain of experimentally and theoretically admissible
parameters that maximize the yield of the
sterile component. This will allow us to give the
final answer to the question of whether StN can affect
the collapse dynamics and parameters. However, the
peculiarities of the StN spectra themselves (namely
their significant hardness at high energies) seem to be of
interest from the viewpoint of detecting the neutrino
signal from supernovae at ground-based facilities.

In conclusion, let us discuss the influence of the
supernova model used. We considered the collapse of
an iron stellar core with a mass of $2M_\odot$, the so-called
``hot'' collapse typical of very massive stars with a
mass $M\gtrsim 25M_\odot$. Lower-mass stars $(M\sim 15M_\odot)$
have an iron core with a mass of $1.4 M_\odot$ experiencing
``cold'' collapse running with lower electron
capture rates and, hence, with smaller neutronization
of matter (see, e.g., Blinnikov and Okun 1988). As a
result, at the prebounce phase $\WID{\theta}{n}$ will lie even farther
from the resonant value, and, hence, the yield of sterile
neutrinos for cold collapse will be suppressed. In
contrast, at the postbounce phase, as we saw, sterile
neutrinos are generated mainly in the regions of
multiple intersections of the degree of neutronization
$\WID{\theta}{n}$ with its resonant value. As a consequence, the
yield of sterile neutrinos must have close values for
both cold and hot collapse. The same is also true
for the characteristics of the emitted active (electron)
neutrinos within the first several tens of milliseconds
after bounce (see Liebend\"{o}rfer et al. (2004) where
the characteristics of the active neutrino emission from collapsing
$13 M_\odot$ and $40 M_\odot$ stars are compared). As
regards the ``thermal'' postbounce phase (hundreds of
milliseconds), since active neutrinos and antineutrinos
of all flavors are emitted here in approximately
equal numbers, investigating the generation of sterile
neutrinos under these conditions requires a special
consideration.

\section*{ACKNOWLEDGMENTS}
This study was supported in part by the Russian
Foundation for Basic Research
(project \No 14-22-03040 ofi-m). A.V. Yudin and D.K. Nadyozhin also
thank the Russian Science Foundation
(grant \No 14-12-00203) for financial support in the part of this work
concerning the calculation of the characteristics of
active neutrinos. We are grateful to the referees for
constructive remarks.

\end{document}